\documentclass[12pt,a4paper]{article}
\usepackage{amsmath}
\usepackage{amsfonts}
\usepackage{amssymb}
\usepackage{color}
\usepackage[section]{placeins}
\usepackage{here}
\usepackage{subcaption}
\usepackage{graphicx}
\usepackage{epstopdf}
\usepackage{hyperref}
\usepackage[left=2cm,top=2cm,bottom=2cm,right=2cm,
head=0.1cm,foot=1.1cm]{geometry}
\begin{document}
\begin{center}
	{\bf \LARGE
		
Eigensolutions and Thermodynamic Properties of Kratzer plus generalized Morse Potential	}
\end{center}

\vspace{4mm}
\begin{center}
 {\Large{\bf Cecilia N. Isonguyo $^{a}$$^{,}$}}\footnote{\scriptsize E-mail:~ ceciliaisonguyo@uniuyo.edu.ng}, {\Large{\bf Ituen B. Okon $^{a}$$^{,}$}}\footnote{\scriptsize E-mail:~ ituenokon@uniuyo.edu.ng}\Large{\bf }, {\Large{\bf  Akaninyene D. Antia $^{a}$$^{,}$}}\footnote{\scriptsize E-mail:~ akaninyeneantia@uniuyo.edu.ng}, {\Large{\bf  Kayode J. Oyewumi $^{b}$$^{,}$}}\footnote{\scriptsize E-mail:~ kjoyewumi66@unilorin.edu.ng},{\Large{\bf  Ekwevugbe Omugbe $^{c}$$^{,}$}}\footnote{\scriptsize E-mail:~ omugbeekwevugbe@gmail.com},{\Large{\bf  Clement A. Onate $^{d}$$^{,}$}}\footnote{\scriptsize E-mail:~ oaclems14@physicist.net}, {\Large{\bf  Roseline U. Joshua $^{e}$$^{,}$}}\footnote{\scriptsize E-mail:~ roselineukemejoshua@gmail.com} \\ 
 {\Large{\bf Monday E. Udoh $^{a}$$^{,}$}}\footnote{\scriptsize E-mail:~ etiufanmonday@gmail.com} and {\Large{\bf Eno E. Ituen $^{a}$$^{,}$}}\footnote{\scriptsize E-mail:~ enoeituen@uniuyo.edu.ng}
\end{center}
{\small
\begin{center}
{\it $^\textbf{a}$Theoretical Physics Group, Department of Physics, University of Uyo, Akwa Ibom State, Nigeria. }\\
{\it $^\textbf{b}$Department of Physics, University of Ilorin, Kwara State, Nigeria.} \\
{\it $^\textbf{c}$  Department of Physics, Federal University of Petroleum Resources, Effurun, Delta State, Nigeria. } \\{\it $^\textbf{d}$ Department of Physical Sciences, Landmark University, Omu-Aran, Nigeria } \\{\it $^\textbf{e}$ Physics Programme, Department of Physics, University of Uyo, Akwa Ibom State, Nigeria. }
\end{center}}
Corresponding Author's  email: ituenokon@uniuyo.edu.ng

\begin{abstract}
\noindent
In this study, we apply the parametric Nikiforov-Uvarov method to obtain the bound state  solution of Schr\"{o}dinger wave equation in the presence of Kratzer plus generalized Morse potential (KPGM). The energy eigen equation and the corresponding normalised wave function were obtained in closed form. The resulting energy eigen equation were use to study partition function and other  thermodynamic properties such as vibrational mean energy, vibrational specific heat capacity, vibrational mean free energy and vibrational entropy for the proposed potential as applied to lithium hydride diatomic molecule. The thermodynamic plots obtained were in excellent agreement to work of existing literatures. The wave function and probability density plots for the diatomic molecules were obtained through a well designed and implemented maple programme.

\end{abstract}



\section{Introduction}
The exact solutions of Schr\"{o}dinger wave equation is one of the essential part in quantum mechanics, this is because Schr\"{o}dinger wave equation is used to describe non-relativistic spinless particles and also has many applications in atomic, nuclear and high energy Physics \cite{Q1,Q2,Q3,Q4,Q5,Q6,Q7,Q8,Q9}. This has prompted many researchers over the  years to search for the solution of Schcr\"{o}dinger wave equation with different potentials \cite{Q10,Q11,Q12,Q13,Q14,Q15}. However, different methods have been used to obtain approximate solution of Schcr\"{o}dinger wave equation, they include Nikiforov-Uvarov method (NU) \cite{Q16,Q17,Q18,Q19,Q20,Q21,Q22}, Supersymmetry quantum mechanics (SUSY) \cite{Q23,Q24,Q25,Q26,Q27,Q28}, Asymtotic Iteration method (AIM)\cite{Q29,Q30}, Factorization method \cite{Q31,Q32}, Exact and proper quantization method \cite{Q33,Q34,Q35,Q36}. In solving the wave equation, the results obtained for various potential models are vastly applied \cite{Q4,Q31,Q37,Q38,Q39,Q38,Q39,Q40,Q41}.  

The thermodynamic properties of a particular system is studied by finding the partition function which is a function of temperature. Other thermodynamic properties such as entropy, specific heat capacity, mean free energy and others are easily obtained using the partition function, which is widely applied in molecular physics and statistical physics \cite{Q42,Q43,Q44}.

In this research article, we solve the bound state solution of Schr\"{o}dinger wave equation with Kratzer plus generalized Morse potential using the parametric Nikiforov-Uvarov method. We also extend our work to study the thermodynamic properties of the system we are studying. The Kratzer plus generalized Morse potential takes the form 

\begin{eqnarray} \label{GrindEQ__1_} 
V(r)= -2D_e\left(\frac{r_e}{r}-\frac{r^2_e}{2r^2}\right) + D\left(1-\frac{be^{\alpha r}}{1-e^{\alpha r}}\right)^2,
\end{eqnarray}
where $D_e$ is dissociation energy, $r_e$is the equilibrium bond length, $r$ represent the interatomic distance, $\alpha$ is the screening parameter. This article is organised as follows: Section 1 is the introduction of the article, The Parametric Nikiforov-Uvarov method and the non-relativistic solution is presented in section 2, Thermodynamic properties is expressed in section 3, while the numerical solution is shown in section 4.

\section{The  Parametric Nikiforov-Uvarov(NU) Method}
 In the parametric NU method, the second order linear differential equation is reduced  to a generalised equation of hyper-geometric type which provides exact solutions interms of special orthogonal functions and the corresponding energy eigenvalues of the form. With the appropriate coordinate transformation $S=S(x)$  the equation can be written as  \cite{Q17,Q45,Q46,Q47,Q48,Q49}.  
\begin{eqnarray} \label{GrindEQ__2a_}
\psi ''(s)+\frac{\tilde{\tau}(s)}{\sigma(s)} \psi'(s)+ \frac{\tilde{\sigma}(s)}{\sigma^2(s)}\psi(s)=0
\end{eqnarray} 
where $\tilde{\tau}(s)$ is the polynomial of degree one, $ \sigma(s) $ and $ \tilde{\sigma}(s)$ are polynomials of at most degree two. Then the parametric NU differential equation is in the form \cite{Q45} $ $ 
\begin{eqnarray} \label{GrindEQ__2_} 
\psi ''(s)+\frac{(c_{1} -c_{2} s)}{s(1-c_{3} s)} \psi '(s)+\frac{1}{s^{2} (1-c_{3} s)^{2} } \left[-\Omega_{1} s^{2} +\Omega_{2} s-\Omega_{3} \right]\psi (s)=0 
\end{eqnarray}
The parametric constants are obtained as follows
\begin{eqnarray} \label{GrindEQ__3_} 
\left. \begin{array}{l} c_{1}=c_{2}=c_{3}=1;\,\, {c_{4} =\frac{1}{2} \left(1-c_{1} \right);\, \, \, c_{5} =\frac{1}{2} \left(c_{2} -c_{3} \right);\, \, c_{6} =c_{5}^{2} +\epsilon _{1} } \\ {c_{7} =2c_{4} c_{5} -\Omega _{2} ;\, \, c_{8} =c_{4}^{2} +\Omega _{3} ;\, \, \, c_{9} =c_{3} c_{7} +c_{3}^{2} c_{8} +c_{6} } \\ {c_{10} =c_{1} +2c_{4} +2\sqrt{c_{8} } ;\, \, \, c_{11} =c_{2} -2c_{5} +2\left(\sqrt{c_{9} } +c_{3} \sqrt{c_{8} } \right)} \\ {c_{12} =c_{4} +\sqrt{c_{8} } ;\, \, c_{13} =c_{5} -\left(\sqrt{c_{9} } +c_{3} \sqrt{c_{8} } \right)} \end{array}\right\} .
\end{eqnarray} 
The eigen energy equation is given as 
\begin{eqnarray} \label{GrindEQ_4__} 
c_{2}^{} n-\left(2n+1\right)c_{5} \left(2n+1\right)\left(\sqrt{c_{9} } +c_{3} \sqrt{c_{8} } \right)+n\left(n-1\right)c_{3} +c_{7} 2c_{3} c_{8} +2\sqrt{c_{8} c_{9} } =0 
\end{eqnarray}
The corresponding total wave function is then given as 
\begin{eqnarray} \label{GrindEQ__5_} 
\Psi (s)=N_{nl} s^{c_{12} } \left(1-c_{3} s\right)^{-c_{12} -\frac{c_{11} }{c_{3} } } P_{n}^{\left(c_{10} -1,\, \, \frac{c_{11} }{c_{3} } -c_{10} -1\right)} \left(1-2c_{3} s\right) 
\end{eqnarray}

\subsection{Non-relativistic solution with KPGMP}
The Schr\"{o}dinger wave equation for an arbitrary external potential ~$V(r)$~in spherical coordinate is written as \cite{Q4}
\begin{eqnarray}\label{GrindEQ__6_}
\frac{d^2 \psi_{n\ell}(r)}{d r^2}+\frac{2\mu}{\hbar^2}\left[E-V(r)-\frac{\hbar^2 \ell(\ell+1)}{2\hbar r^2}\right]\psi_{n\ell}= 0
\end{eqnarray}
where $E$ is the exact bound state energy eigenvalues, ~$R_{n\ell} (r)$~ is the eigenfunction, $\mu=\frac{m_1m_2}{m_1+m_2}$ being the reduced mass, ~$(\hbar=\mu=1)$. $n$ denotes the principal quantum number ($n$ and $\ell$ are known as the vibration-rotation quantum numbers), $r$ is the internuclear separation. 

Also, on substituting equation (\ref{GrindEQ__1_}) into equation (\ref{GrindEQ__6_}), the radial part of the Schr\"{o}dinger equation for the KPGM is given as 
\begin{eqnarray}\label{GrindEQ__7_}
\frac{d^2 \psi_{n\ell} (r)}{d r^2} + \frac{2\mu}{\hbar^2}\left[ E+ \left(-2D_e\left(\frac{r_e}{r}-\frac{r^2_e}{2r^2}\right) + D\left(1-\frac{be^{\alpha r}}{1-e^{\alpha r}}\right)^2\right)-\frac{ \hbar^2 \ell (\ell+1)}{ 2 \mu r^2}\right]\psi_{n\ell} (r)=0.
\label{E10}
\end{eqnarray} 
The Green-Aldrich approximation is given as \cite{Q50}
\begin{eqnarray}\label{GrindEQ__8_}
\frac{1}{r^2} = \frac{4\alpha^2 e^{-2\alpha r}}{(1-e^{-\alpha r})^2} ~~,~~ \frac{1}{r} = \frac{2\alpha e^{-\alpha r}}{1-e^{-\alpha r}}.
\end{eqnarray}

On Substituting the transformation $s=e^{\alpha r}$ and applying the Green-Aldrich approximation into equation (\ref{GrindEQ__7_}) yields

\begin{eqnarray}\label{GrindEQ__9_}
\frac{d^2 \psi(s)}{dr^2}+\frac{(1-s)}{s(1-s)}\frac{d\psi(s)}{dr}+ \frac{1}{s^2(1-2)^2}\left[ \begin{array}{l} (-\xi^2 -A -B -C - F -G -4\lambda)s^2 \\ + (2\xi^2 +A +2C +F)s -(-\xi^2-C) \end{array} \right]\psi(s) =0 
\end{eqnarray}
where
\begin{eqnarray}\label{GrindEQ__10_}
-\xi^2=\frac{2\mu E}{\alpha^2 \hbar^2} ~, A=\frac{8 \mu D_e r_{e}}{\alpha \hbar^2} ~, B= \frac{8 \mu D_e r^2_{e}}{\alpha^2 \hbar^2} ~, C= \frac{2\mu D}{\alpha^2 \hbar^2}~, F\frac{4\mu D b}{\alpha^2 \hbar^2} ~, G= \frac{2\mu D b^2}{\alpha^2 \hbar^2}.
\end{eqnarray}
Comparing equation (\ref{GrindEQ__9_}) with the standard parametric NU differential equation of (\ref{GrindEQ__2_}), the parameters are obtained as follows:

\begin{eqnarray}\label{GrindEQ__11_}
\left. \begin{array}{lll} 
\Omega_1 &=&\xi^2 -A -B -C - F -G -4\lambda, \Omega_2= 2\xi^2 +A +2C +F, ~~ \Omega_3= \xi^2-C   \\ 
c_1 &=& c_2=c_3=1,~ c_4=0,~ c_5=-\frac{1}{2},~ c_6= \frac{1}{4}+\xi +A +B +C +F +G +4\lambda, \\
c_7 &=& -2\xi^2-A-2C-F,~ c_8=\xi^2,~ c_9= \frac{1}{4}+B +G +4\lambda,~ c_10=1+2 \sqrt{\xi^2 +C} \\
c_{11} &=& 2+2\left(\sqrt{\frac{1}{4}+B +G +C +4\lambda+\xi^2}\right),~ c_{12}=\sqrt{\xi^2 +C}~, \\
c_{13} &=&  -\frac{1}{2} -\left(\sqrt{\frac{1}{4}+B +G +C +4\lambda+\xi^2}\right)  \end{array} \right\}.
\end{eqnarray}

By substituting the appropriate parameters of equation (\ref{GrindEQ__11_}) into equations (\ref{GrindEQ__5_}) and equation (\ref{GrindEQ_4__}), then simplify gives 
the respective wave function  and energy eigenvalue equation for the KPGMP as
\begin{eqnarray}\label{GrindEQ__12_}
\psi_{n\ell}= N_{n\ell}s^{\gamma}\left(1-s\right)^{\delta} P_{n}^{\left[2\gamma,2\delta-1\right]}\left(1-2s\right) ,\ \ \ s=e^{\alpha r}
\end{eqnarray}
and
\begin{eqnarray}\label{GrindEQ__13_} 
E_{n\ell}=-\frac{\alpha^{2}\hbar^{2}}{2\mu }\left[\frac{\left(n^2+n+\frac{1}{2}\right)+(2n+1)\eta-\frac{8\mu D_e r_e}{\alpha \hbar^2}-\frac{8\mu D b}{\alpha^{2}\hbar^{2}}}{(2n+1) +2\eta}\right]^{2}-\frac{2\mu D b^2}{\alpha^{2} \hbar^2}
\end{eqnarray}
where
\begin{eqnarray}\label{GrindEQ__14_}
\gamma=\sqrt{\frac{2\mu D}{\alpha^{2} \hbar^2}-\frac{2\mu E}{\alpha^{2} \hbar^2}}, \ \ \delta= \frac{1}{2}+\sqrt{\frac{1}{4} + \frac{8\mu D_e r_e^{2}}{\alpha^{2} \hbar^2}+ \frac{2\mu D b^2}{\alpha^{2} \hbar^2}+ 4\ell(\ell+1)} , \notag\\
\eta=\sqrt{\frac{1}{4}+ \frac{8\mu D_e r_e^{2}}{\alpha^{2} \hbar^2}+ \frac{2\mu D b^2}{\alpha^{2} \hbar^2}+ 4\ell(\ell+1)}
\end{eqnarray}
The normalization constant in equation \ref{GrindEQ__12_} can be obtain using the normalization condition \cite{Q4,Q1}
\begin{eqnarray}\label{GrindEQ__15_}
\int_{0}^{\infty} |\psi_{n\ell}|^2 dr=\int_{0}^{\infty} |N_{n\ell}s^{\gamma}\left(1-s\right)^{\delta} P_{n}^{\left[2\gamma,2\delta-1\right]}\left(1-2s\right)|^2 ds =1 ,
\end{eqnarray}
the wavefunction is assumed to be in the bound at $r \ \varepsilon \ (0,\infty)$ and $ s=e^{\alpha r} \ \varepsilon \ (1,0)$ eqn \ref{GrindEQ__15_} becomes
\begin{eqnarray}\label{GrindEQ__16_}
-\frac{N^{2}_{n\ell}}{\alpha}\int_{0}^{1} s^{2\gamma}\left(1-s\right)^{2\delta} P_{n}^{\left[2\gamma,2\delta-1\right]}\left(1-2s\right)|^2 \frac{ds}{s}  =1
.\end{eqnarray}
Let $ z=1-2s $ thus, the limit of integration of eqn\ref{GrindEQ__16_} changes from $ s \ \varepsilon (1,0)$ to $ z \ \varepsilon (-1,1) $. Then equation \ref{GrindEQ__16_} reduces to
\begin{eqnarray}\label{GrindEQ__17_}
\frac{N^{2}_{n\ell}}{2\alpha}\int_{-1}^{1}\left(\frac{1-z}{2}\right)^{2\gamma-1} \left(\frac{1+z}{2}\right)^{2\delta} \left[P_{n}^{\left[2\gamma,2\delta-1\right]} (z) \right]^2 dz =1.
\end{eqnarray}
Applying the standard integral \cite{Q1,Q4}
\begin{eqnarray}\label{GrindEQ__18_}
\int_{-1}^{1}\left(\frac{1-w}{2}\right)^x \left(\frac{1+w}{2}\right)^y \left[P_{n}^{(x,y-1)} (w) \right]^2 dw =\frac{2^{x+y+1} \Gamma(x+n+1) \Gamma(y+n+1) }{n !\Gamma(x+y+n+1) \Gamma(x+y+2n+1) }.
\end{eqnarray}
Also, let $ z=w,\ x=2\gamma-1,\ y=2\delta $. Then the normalization constant can be obtained as 
\begin{eqnarray}\label{GrindEQ__19_}
N_{n\ell}=\sqrt{\frac{2\alpha (n!) \Gamma(2\gamma+2\delta+n)\Gamma(2\gamma+2\delta+2n)}{2^{(2\gamma+2\delta)}\Gamma(2\gamma+n)\Gamma(2\delta+n+1)}} ,
\end{eqnarray}
therefore, the total normalized wave function is given as
\begin{eqnarray}\label{GrindEQ__20_}
\psi (s)=\sqrt{\frac{2\alpha (n!) \Gamma(2\gamma+2\delta+n)\Gamma(2\gamma+2\delta+2n)}{2^{(2\gamma+2\delta)}\Gamma(2\gamma+n)\Gamma(2\delta+n+1)}} \ \ s^{\gamma}(1-s)^{\delta} P_{n}^{\left[2\gamma,2\delta-1\right]} (1-2s).
\end{eqnarray}

\begin{figure}[H]
	\begin{subfigure}[b]{0.5\textwidth}
		\includegraphics[width=9cm, height=9cm]{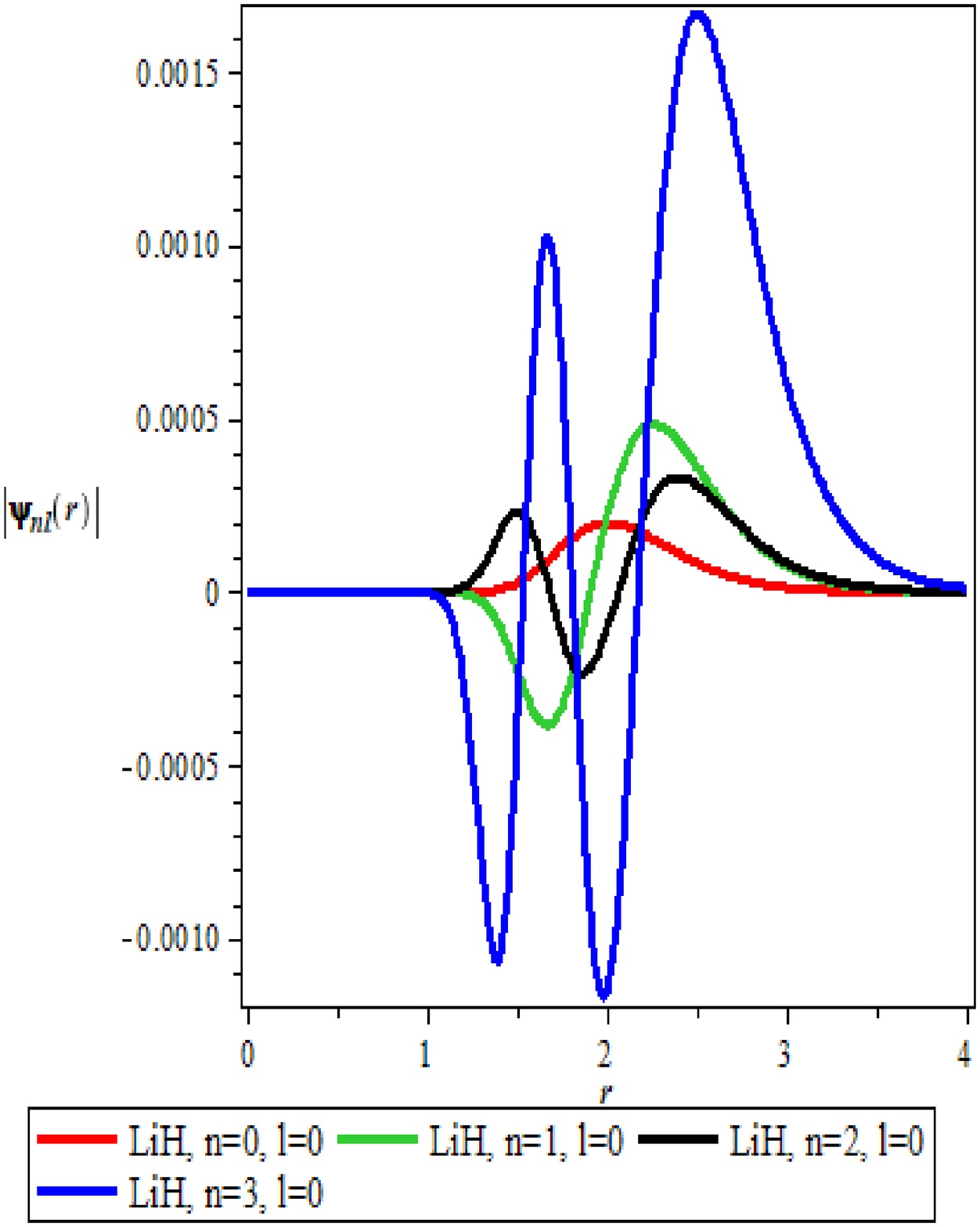}
		\caption*{Figure 1(a): Wavefunction plot for fixed $l=0$ for LiH molecule }
	\end{subfigure}\hspace{1cm} 
	~ 
	\begin{subfigure}[b]{0.5\textwidth}
		\includegraphics[width=9cm, height=9cm]{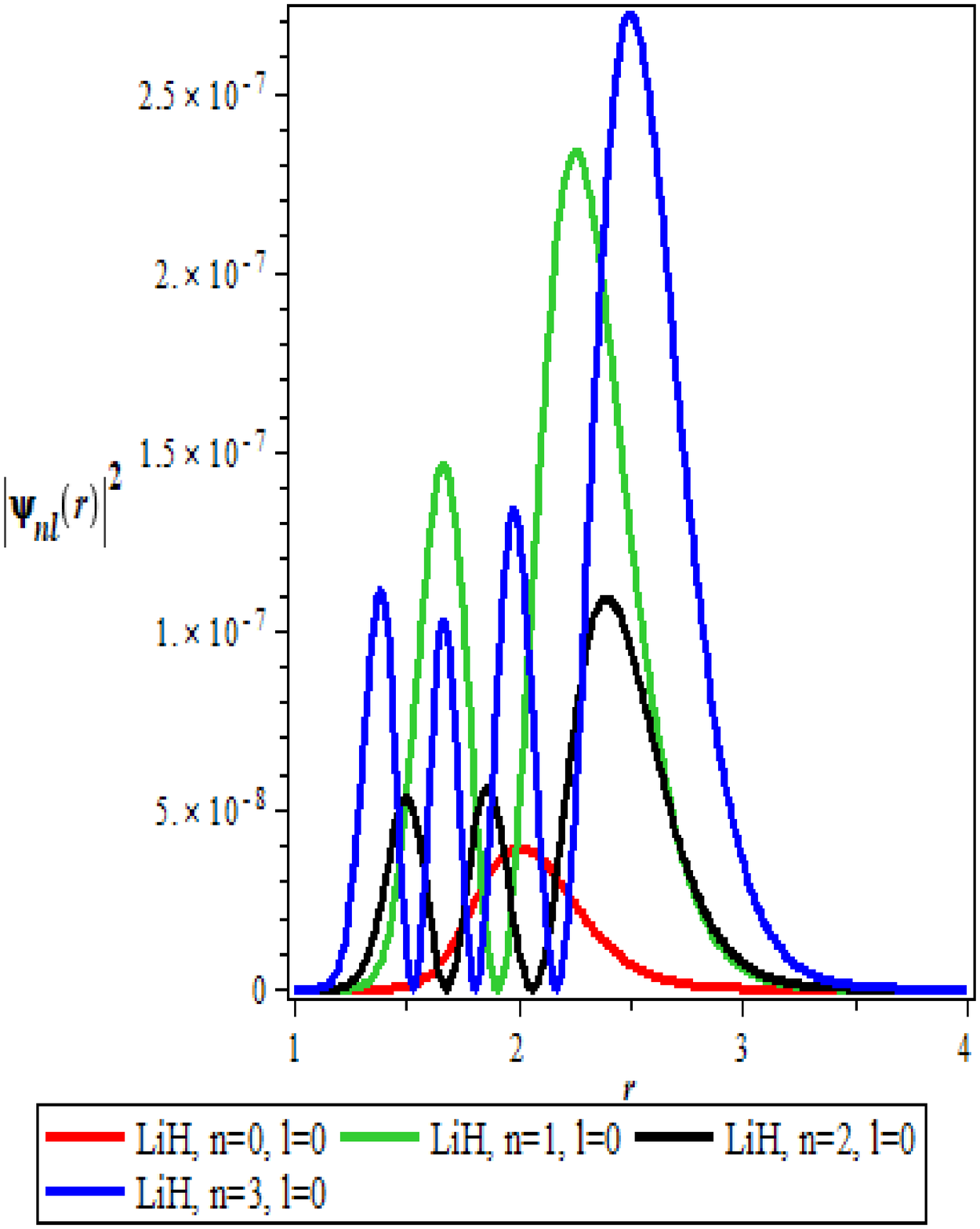}
		\caption*{Figure 1(b): Probabilty density plot for fixed $l=0$ for LiH molecule}
	\end{subfigure}
	~ 
	\begin{subfigure}[b]{0.5\textwidth}
		\includegraphics[width=9cm, height=9cm]{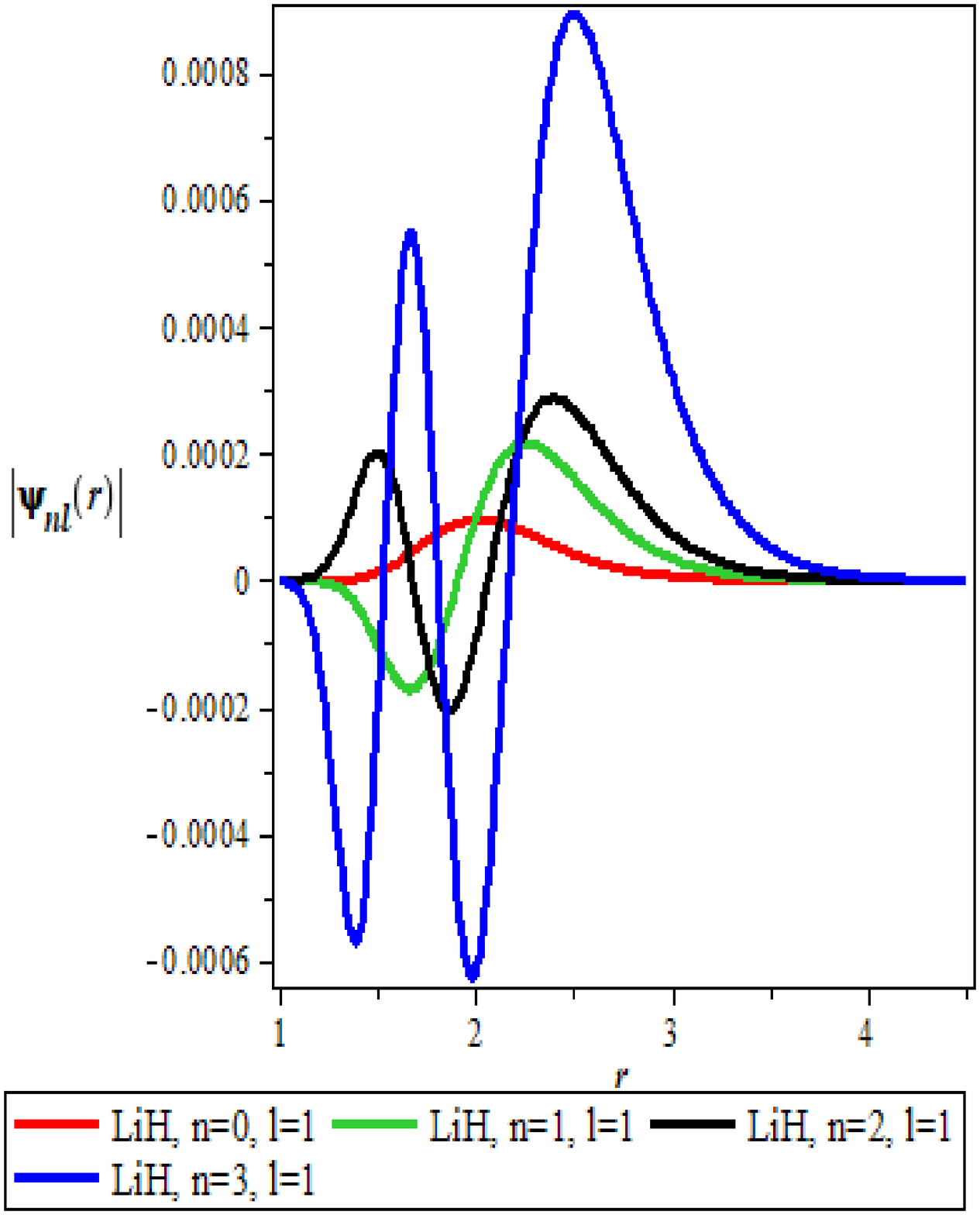}
		\caption*{Figure 2(a): Wavefunction plot for fixed $l=1$ for LiH molecule}
	\end{subfigure}\hspace{1cm}
	\begin{subfigure}[b]{0.5\textwidth}
		\includegraphics[width=9cm, height=9cm]{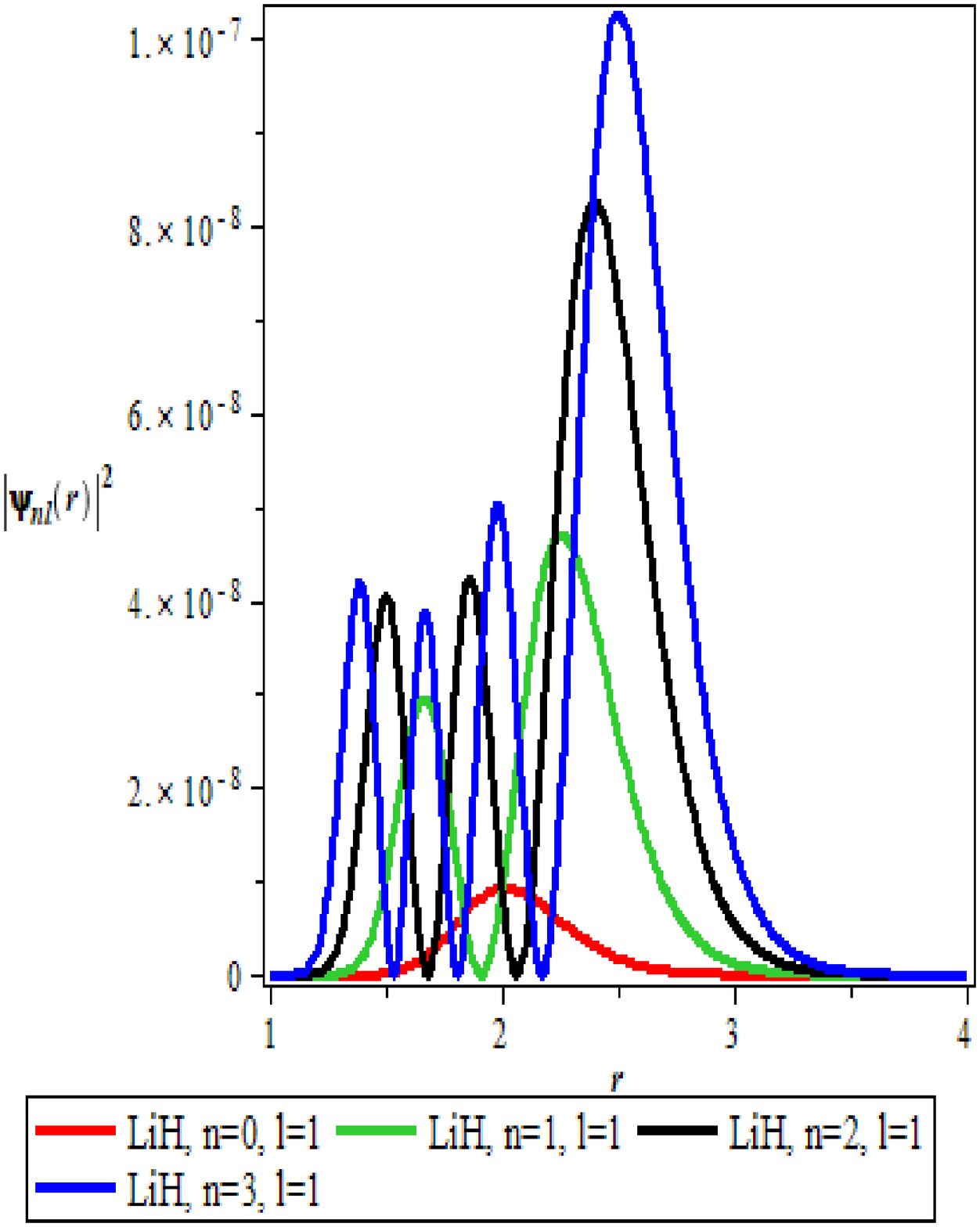}
		\caption*{Figure 2(b): Probabilty density plot for fixed $l=1$ for LiH molecule}
	\end{subfigure}
\end{figure}
\begin{figure}
	\centering
	\begin{subfigure}[b]{0.5\textwidth}
		\includegraphics[width=9cm, height=9cm]{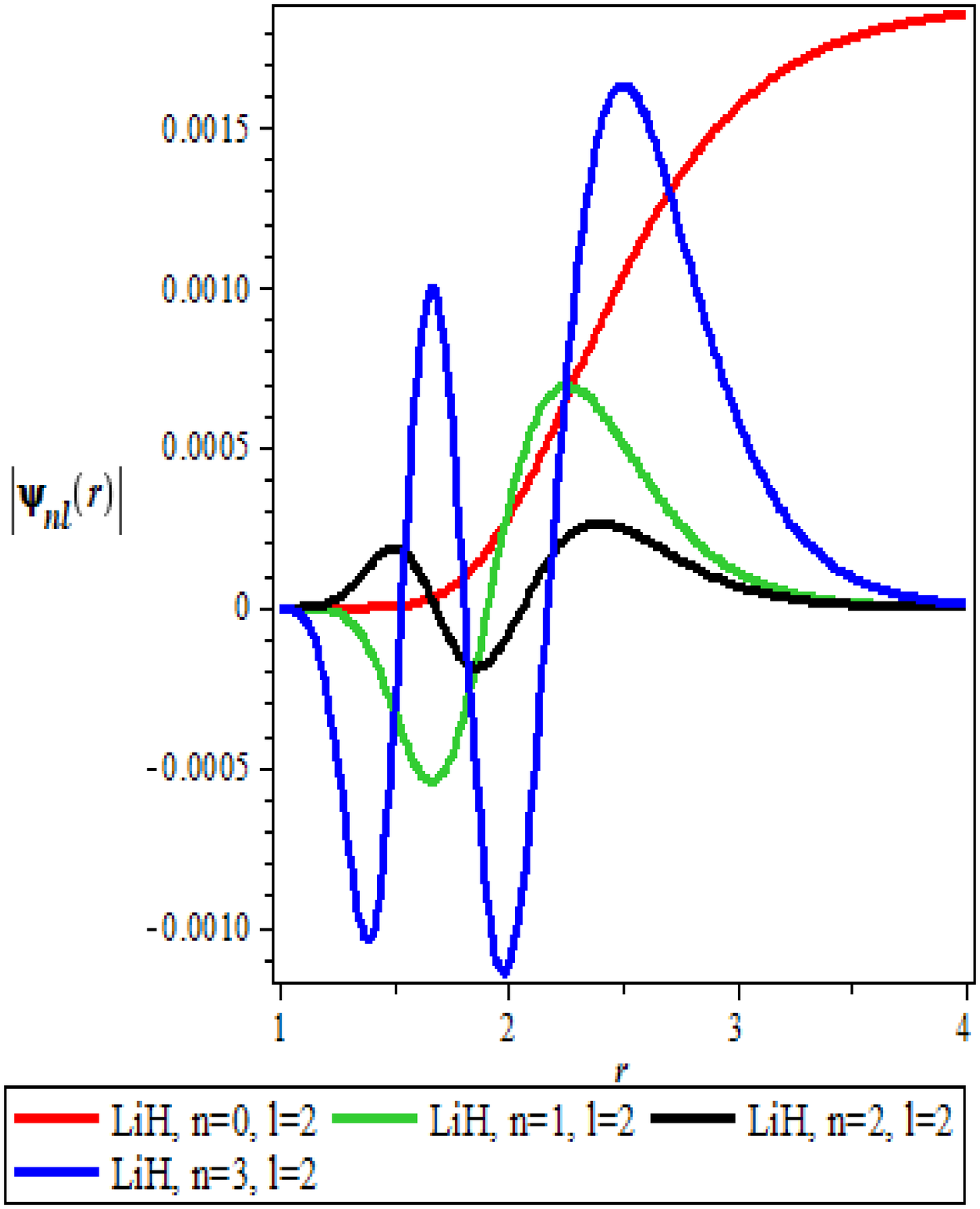}
		\caption*{Figure 3(a): Wavefunction plot for fixed $l=2$ \\ for LiH molecule}
	\end{subfigure}
	~ 
	\begin{subfigure}[b]{0.5\textwidth}
		\includegraphics[width=9cm, height=9cm]{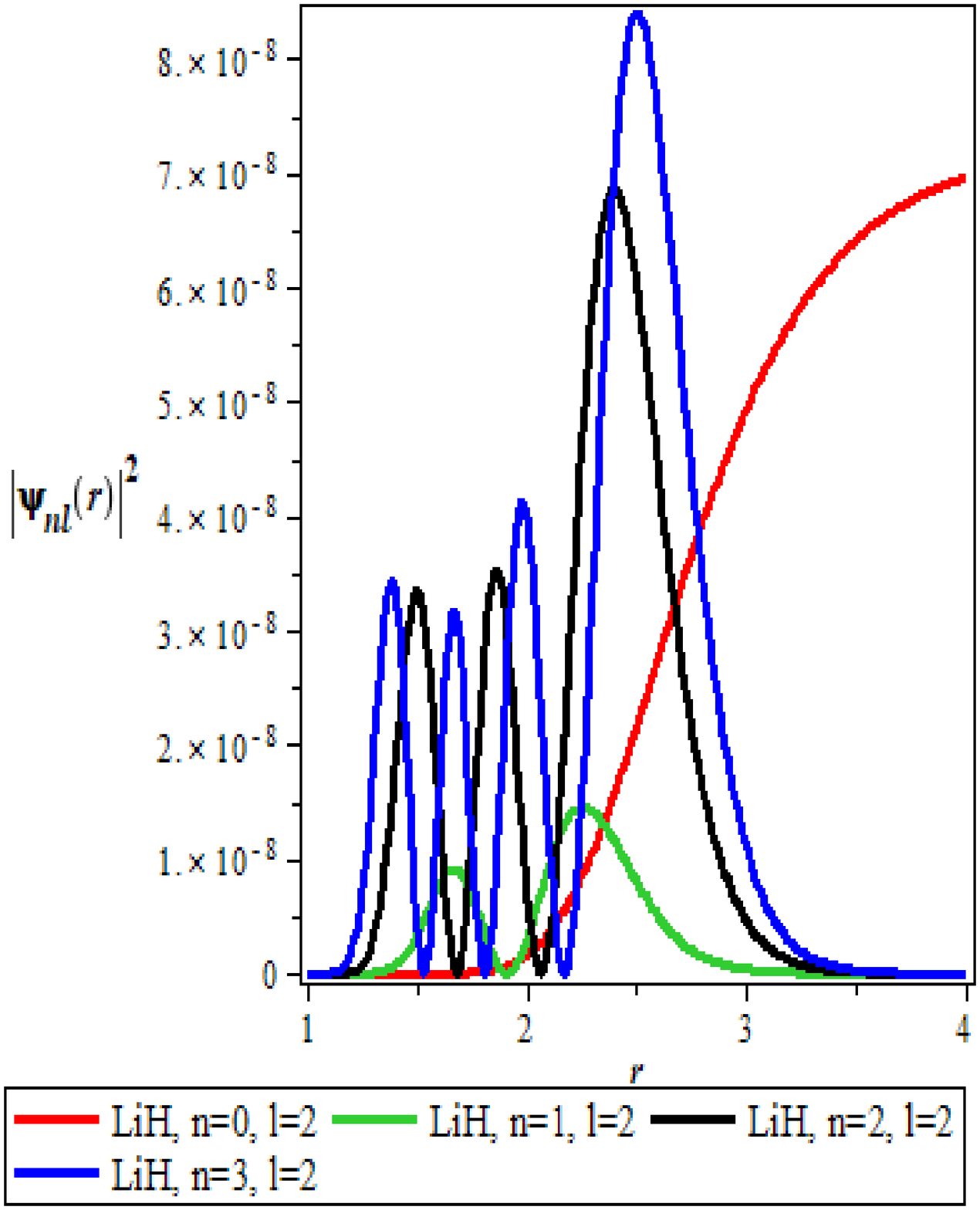}
		\caption*{Figure 3(b): Probabilty density plot for fixed $l=2$ for LiH molecule}
	\end{subfigure}
	~ 
	
\end{figure}
\begin{figure}
	\centering
	\begin{subfigure}[b]{0.5\textwidth}
		\includegraphics[width=9cm, height=9cm]{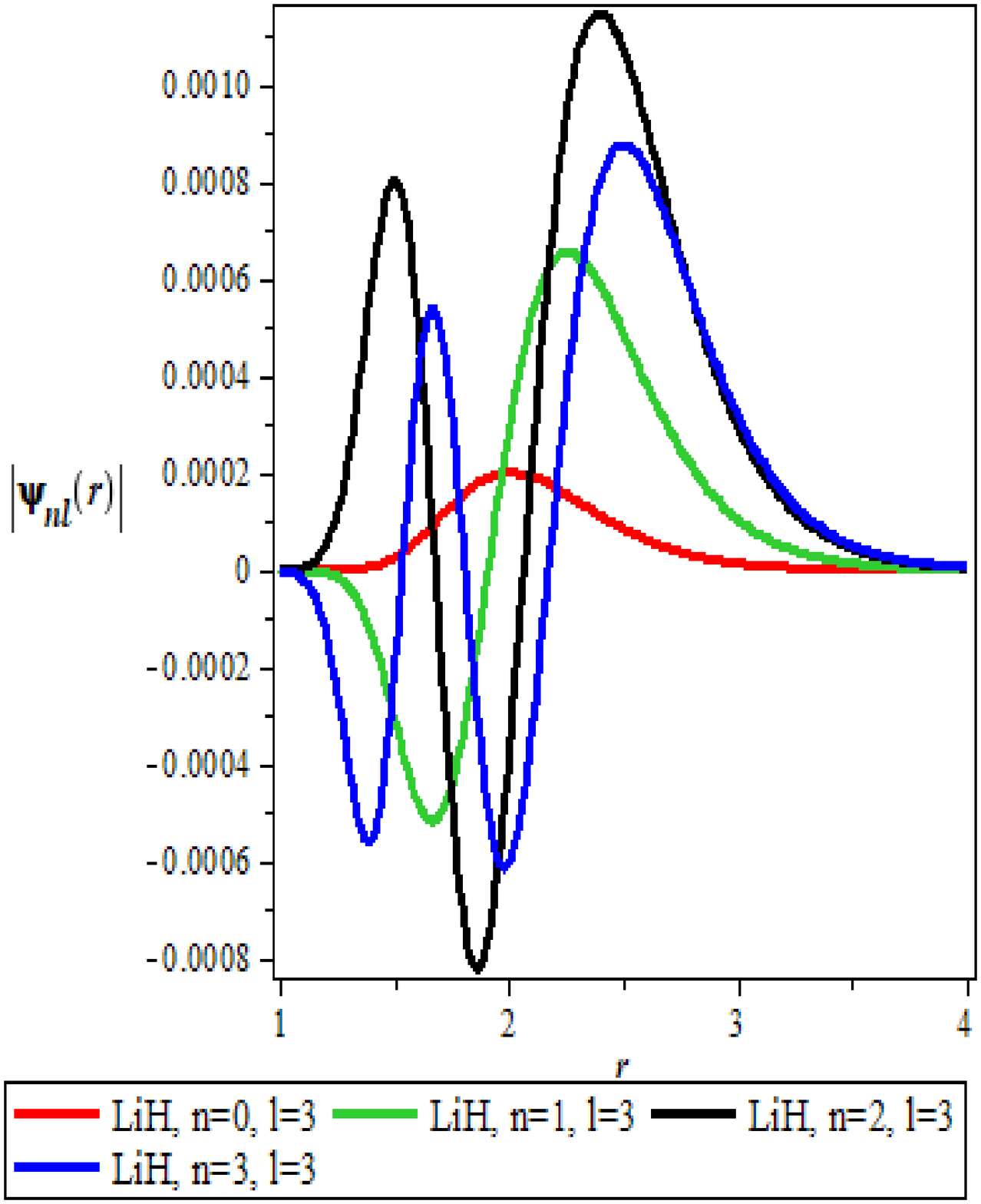}
		\caption*{Figure 4(a):  Wavefunction plot for fixed $l=3$ \\ for LiH molecule}
	\end{subfigure}
	~ 
	\begin{subfigure}[b]{0.5\textwidth}
		\includegraphics[width=9cm, height=9cm]{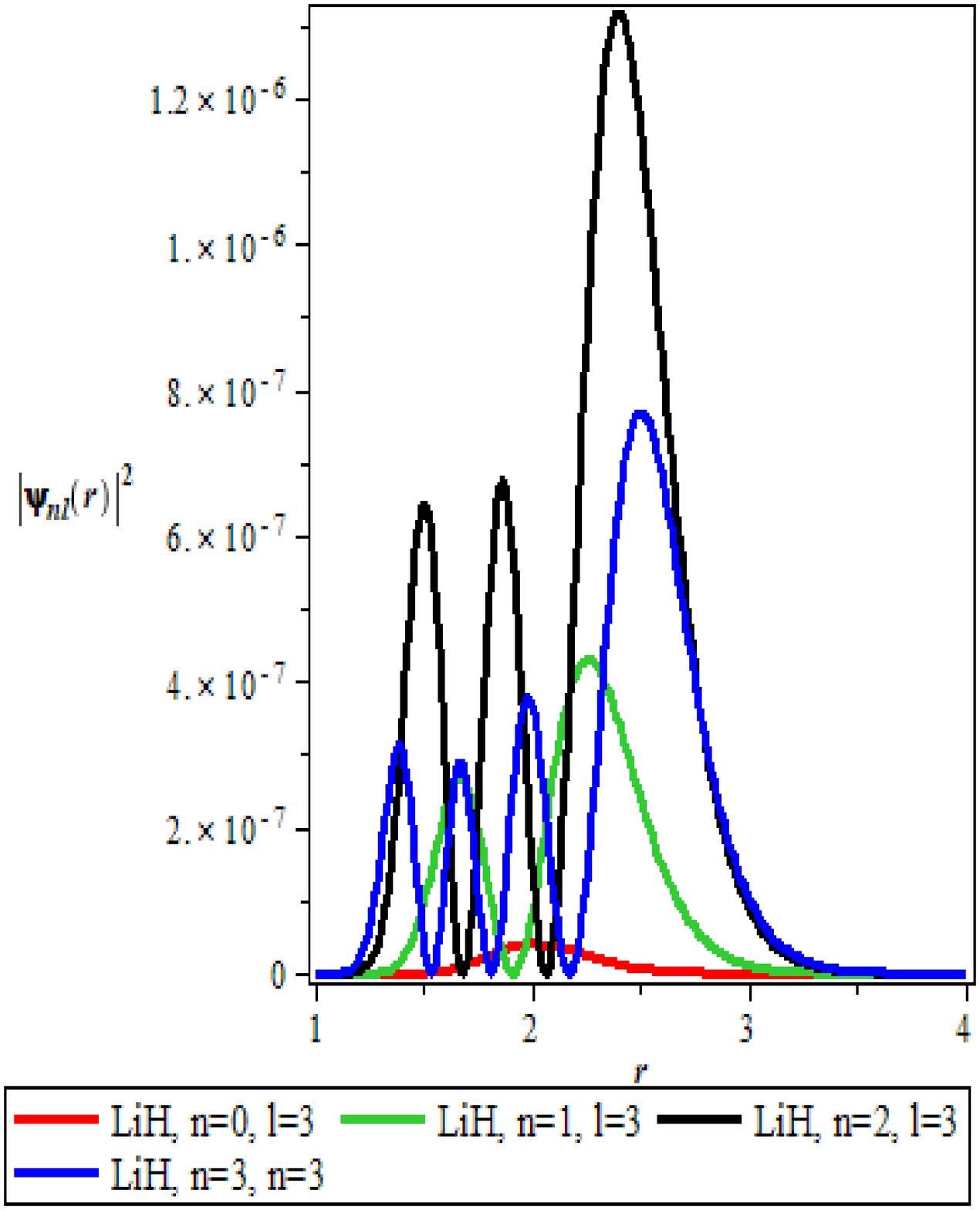}
		\caption*{Figure 4(b): Probabilty density plot for fixed $l=3$  for LiH molecule}
	\end{subfigure}
	~ 
	
\end{figure}

\section{Thermodynamic properties for the Potential model}

The thermodynamic properties for KPGM will be studied by first obtaining the vibrational partition function defined as 
\begin{eqnarray}\label{GrindEQ__21_} 
Z(\beta)=\sum_{n=0}^{\lambda} e^{-\beta E_{n}}
\end{eqnarray}
where $\lambda$ is an upper bound of the vibrational quantum number obtain from the numerical solution of $\frac{dE_{n}}{dn}=0$,  $\beta =\frac{1}{kT}$ where $K$ and $T$are Boltzmann constant and absolute temperature respectively. In the classical limit, the summation in \ref{GrindEQ__21_} can be replaced with the integral  :

\begin{eqnarray}\label{GrindEQ__22_} 
Z(\beta) = \int_{0}^{\lambda} e^{-\beta E_{n}}dn
\end{eqnarray}
The energy equation of equation\ref{GrindEQ__13_} can be simplified to

\begin{eqnarray}\label{GrindEQ__23_} 
E_{nl}=Q_{1}-Q_{2}\left[(n+\Delta)+\frac{Q_{3}}{(n+\Delta)}\right]^{2}
\end{eqnarray}
where 
\begin{eqnarray}\label{GrindEQ__24_} 
Q_{1}=\frac{-2\mu D b^2}{\alpha^{2} \hbar^{2} }, Q_{2}=\frac{\alpha^{2} \hbar^{2}}{8\mu}, Q_{3}=-\frac{\mu D_e r_e^{2} }{2 \alpha^{2} \hbar^{2} }-\frac{\mu D b^2}{8\alpha^{2} \hbar^{2}}-\frac{\ell(\ell+1)}{4}-\frac{4\mu D_e }{\alpha^{2} \hbar^{2}},\,\   \notag \\ 
 \Delta=\frac{1}{2}+\frac{1}{2}\sqrt{1+\frac{2\mu D_e r_e^{2}}{\alpha^{2} \hbar^{2}}+\frac{\mu D b^2}{2\alpha^{2} \hbar^{2}}+\ell(\ell+1)}
\end{eqnarray}
 The energy equation \ref{GrindEQ__23_} can then be express in the form 
\begin{eqnarray}\label{GrindEQ__25_} 
E_{nl}=-\left[Q_{2}\rho^{2}+\frac{Q_{2}Q_{3}^{2}}{\rho^{2}}\right]-[2Q_2 Q_3 -Q_1 ] ,\,\, \rho = n+\Delta
\end{eqnarray}
Hence, the partition function equation \ref{GrindEQ__22_} can be express in the classical limit as
\begin{eqnarray}\label{GrindEQ__26_} 
Z(\beta)=e^{\beta(2 Q_{2} Q_{3}-Q_1)}\int_{0}^{\lambda}e^{(Q_{2}\rho^{2}+\frac{Q_{2}Q_{3}^{2}}{\rho^{2}})}d\rho
\end{eqnarray}
Equation\ref{GrindEQ__22_} is integrated using MAPLE package. Hence, the integral equation \ref{GrindEQ__22_}  which is the partition function is given as 

\begin{eqnarray}\label{GrindEQ__27_} 
Z(\beta)=\frac{\zeta_1}{4 \sqrt{-\beta  Q_2}} \left[ \begin{array}{l} 1+ \text{erf}\left(\lambda  \sqrt{-\beta  Q_2}-\frac{\sqrt{-\beta  Q_2
		Q_3^2}}{\lambda }\right)-e^{4 \sqrt{-\beta  Q_2} \sqrt{-\beta  Q_2
		Q_3^2}}
	 \\  \text{erfc}\left(\lambda  \sqrt{-\beta  Q_2}+\frac{\sqrt{-\beta  Q_2
	 		Q_3^2}}{\lambda }\right) \end{array} \right]
\end{eqnarray}
where
\begin{eqnarray}\label{GrindEQ__28_} 
\zeta_1=\sqrt{\pi } \exp \left(-\beta  Q_1+2 \beta  Q_2 Q_3-2 \sqrt{-\beta 
	Q_2} \sqrt{-\beta  Q_2 Q_3^2}\right)
\end{eqnarray}
Using the partition function \ref{GrindEQ__27_}, other thermodynamic properties are obtain as follows\\
(a)Vibrational mean energy:

\begin{eqnarray}\label{GrindEQ__29_}  
U(\beta)=-\frac{\partial \ln Z(\beta)}{\partial \beta}= \left[\frac{\begin{array}{l} \sqrt{-\beta  Q_2 Q_3^2} \left(\sqrt{\pi } \zeta_{2} \left(2 \beta  Q_1+1\right)+\zeta_{4}\right) \\
	+\zeta_{2} \sqrt{-\beta  Q_2 Q_3^2}-4 \sqrt{\pi } \beta  \zeta_{3} Q_2 Q_3^2
	\sqrt{-\beta  Q_2} \end{array}}{2 \sqrt{\pi } \beta  \zeta_{2} \sqrt{-\beta  Q_2 Q_3^2}}\right]
\end{eqnarray}
where 
\begin{eqnarray}\label{GrindEQ__30_}
\left. \begin{array}{l} \zeta_{2}=\text{erfc}\left(\lambda  \sqrt{-\beta  Q_2}-\frac{\sqrt{-\beta  Q_2 Q_3^2}}{\lambda }\right)+e^{4 \sqrt{-\beta  Q_2} \sqrt{-\beta  Q_2 Q_3^2}} \text{erfc}\left(\lambda 
\sqrt{-\beta  Q_2}+\frac{\sqrt{-\beta  Q_2 Q_3^2}}{\lambda }\right)-2  \\ 
\zeta_{3} =\text{erfc}\left(\lambda  \sqrt{-\beta  Q_2}-\frac{\sqrt{-\beta  Q_2 Q_3^2}}{\lambda }\right)-e^{4 \sqrt{-\beta  Q_2} \sqrt{-\beta  Q_2 Q_3^2}} \text{erfc}\left(\lambda 
\sqrt{-\beta  Q_2}+\frac{\sqrt{-\beta  Q_2 Q_3^2}}{\lambda }\right)-2 \\ \zeta_{4}= 4 \lambda  \sqrt{-\beta  Q_2} \exp \left(\frac{\beta  Q_2 \left(\lambda ^4+Q_3^2\right)}{\lambda ^2}+2 \sqrt{-\beta  Q_2} \sqrt{-\beta  Q_2 Q_3^2}\right)  \\    \end{array}\right\}.
\end{eqnarray}
(b) Vibrational specific heat capacity:

\begin{eqnarray}\label{GrindEQ__31_}
C(\beta)=k\beta^{2}(\frac{\partial^{2}\ln Z(\beta)}{\partial\beta^{2}})=
\frac{\text{k$\beta $}^2 \left(-\pi  \lambda \zeta_{2}^2 \sqrt{-\beta  Q_2} \sqrt{-\beta  Q_2
		Q_3^2}+\zeta_{8}-\zeta_{9}\right)}{2 \pi  \beta ^2 \lambda  \zeta_{2}^2 \sqrt{-\beta  Q_2} \sqrt{-\beta  Q_2 Q_3^2}} 
\end{eqnarray}

where
\begin{eqnarray}\label{GrindEQ__32_} 
\left. \begin{array}{lll} 
\zeta_{5} &=& e^{\frac{\beta  Q_2 \left(\lambda ^4+Q_3^2\right)}{\lambda ^2}} \left(4 \lambda ^2 \sqrt{-\beta  Q_2}+\sqrt{-\beta  Q_2
	Q_3^2}\right)-8 \sqrt{\pi } \lambda  e^{2 \sqrt{-\beta  Q_2} \sqrt{-\beta  Q_2 Q_3^2}} \\&& \sqrt{-\beta  Q_2} \sqrt{-\beta  Q_2
	Q_3^2}  \text{erfc}\left(\lambda  \sqrt{-\beta  Q_2}+\frac{\sqrt{-\beta  Q_2 Q_3^2}}{\lambda }\right)  \\ 
\zeta_{6} &=& \left(\sqrt{-\beta  Q_2 Q_3^2}-4 \lambda ^2 \sqrt{-\beta  Q_2}\right) \exp \left(\frac{\beta  Q_2 \left(\lambda
	^4+Q_3^2\right)}{\lambda ^2}+2 \sqrt{-\beta  Q_2} \sqrt{-\beta  Q_2 Q_3^2}\right) \\&& +16 \sqrt{\pi } \lambda  \sqrt{-\beta  Q_2}
\sqrt{-\beta  Q_2 Q_3^2} \\ 
\zeta_{7} &=& \zeta_{5} \text{erfc}\left(\lambda  \sqrt{-\beta  Q_2}-\frac{\sqrt{-\beta  Q_2 Q_3^2}}{\lambda }\right)+\zeta_{6} e^{2
	\sqrt{-\beta  Q_2} \sqrt{-\beta  Q_2 Q_3^2}} \\&& \text{erfc}\left(\lambda  \sqrt{-\beta  Q_2}+\frac{\sqrt{-\beta  Q_2
		Q_3^2}}{\lambda }\right)  -2 e^{\frac{\beta  Q_2 \left(\lambda ^4+Q_3^2\right)}{\lambda ^2}} \left(4 \lambda ^2 \sqrt{-\beta 
	Q_2}+\sqrt{-\beta  Q_2 Q_3^2}\right) \\ 
\zeta_{8} &=& 4 \sqrt{\pi } \beta ^2 Q_2^2 e^{2 \sqrt{-\beta  Q_2} \sqrt{-\beta  Q_2 Q_3^2}} \left(\zeta_{7} Q_3^2-\lambda ^4 \zeta_{2}
\sqrt{-\beta  Q_2 Q_3^2} e^{\frac{\beta  Q_2 \left(\lambda ^4+Q_3^2\right)}{\lambda ^2}}\right) \\
\zeta_{9} &=& 2 \beta  \lambda ^2 Q_2 \sqrt{-\beta  Q_2 Q_3^2} \exp \left(\frac{\beta  Q_2 \left(\lambda ^4+Q_3^2\right)}{\lambda ^2}+2
\sqrt{-\beta  Q_2} \sqrt{-\beta  Q_2 Q_3^2}\right) \\&& \left(4 \lambda  \sqrt{-\beta  Q_2} \exp \left(\frac{\beta  Q_2
	\left(\lambda ^4+Q_3^2\right)}{\lambda ^2}+2 \sqrt{-\beta  Q_2} \sqrt{-\beta  Q_2 Q_3^2}\right)-\sqrt{\pi } \zeta_{2}\right) \end{array} \right\}.
\end{eqnarray}

(c) Vibrational entropy
\begin{eqnarray}\label{GrindEQ__33_}
S(\beta) &=& k\ln Z(\beta) - k\beta\frac{\partial \ln Z(\beta)}{\partial\beta}  \notag \\ 
&=&\text{kLog}\left(-\frac{\sqrt{\pi } \zeta_{265} \exp \left(-\beta  Q_1+2 \beta  Q_2 Q_3-2 \sqrt{-\beta  Q_2} \sqrt{-\beta  Q_2
		Q_3^2}\right)}{4 \sqrt{-\beta  Q_2}}\right) \notag \\&& 
+\frac{k \left(\zeta_{10}-\zeta_{11}+\sqrt{\pi } \zeta_{2} \left(2 \beta  Q_1-4 \beta  Q_2 Q_3+1\right)\right)}{2 \sqrt{\pi }
	\zeta_{3}}
\end{eqnarray}
where
\begin{eqnarray}\label{GrindEQ__34_} 
\left. \begin{array}{lll} \zeta_{10} &=& 4 \sqrt{\pi } \sqrt{-\beta  Q_2} \sqrt{-\beta  Q_2 Q_3^2} \text{erfc}\left(\lambda  \sqrt{-\beta  Q_2}-\frac{\sqrt{-\beta  Q_2
		Q_3^2}}{\lambda }\right)  +4 \lambda  \sqrt{-\beta  Q_2} \\&& \exp \left(\frac{\beta  Q_2 \left(\lambda ^4+Q_3^2\right)}{\lambda
	^2}+2 \sqrt{-\beta  Q_2} \sqrt{-\beta  Q_2 Q_3^2}\right)-8 \sqrt{\pi } \sqrt{-\beta  Q_2} \sqrt{-\beta  Q_2 Q_3^2}  \\ 
\zeta_{11} &=& 4 \sqrt{\pi } e^{4 \sqrt{-\beta  Q_2} \sqrt{-\beta  Q_2 Q_3^2}} \sqrt{-\beta  Q_2} \sqrt{-\beta  Q_2 Q_3^2}
\text{erfc}\left(\lambda  \sqrt{-\beta  Q_2}+\frac{\sqrt{-\beta  Q_2 Q_3^2}}{\lambda }\right)    \end{array}\right\}.
\end{eqnarray}
(c) Vibrational free energy
\begin{eqnarray}\label{GrindEQ__35_}
F(\beta) &=& -kT\ln Z(\beta) = \frac{\sqrt{\pi } \zeta_{2} \ln  \exp \left(-\beta  Q_1+2 \beta  Q_2 Q_3-2 \sqrt{-\beta  Q_2} \sqrt{-\beta  Q_2
		Q_3^2}\right)}{4 \beta  \sqrt{-\beta  Q_2}}
\end{eqnarray}

\section{Numerical Results}
\begin{figure}[H]
	\begin{subfigure}[b]{0.5\textwidth}
		\includegraphics[width=9cm, height=9cm]{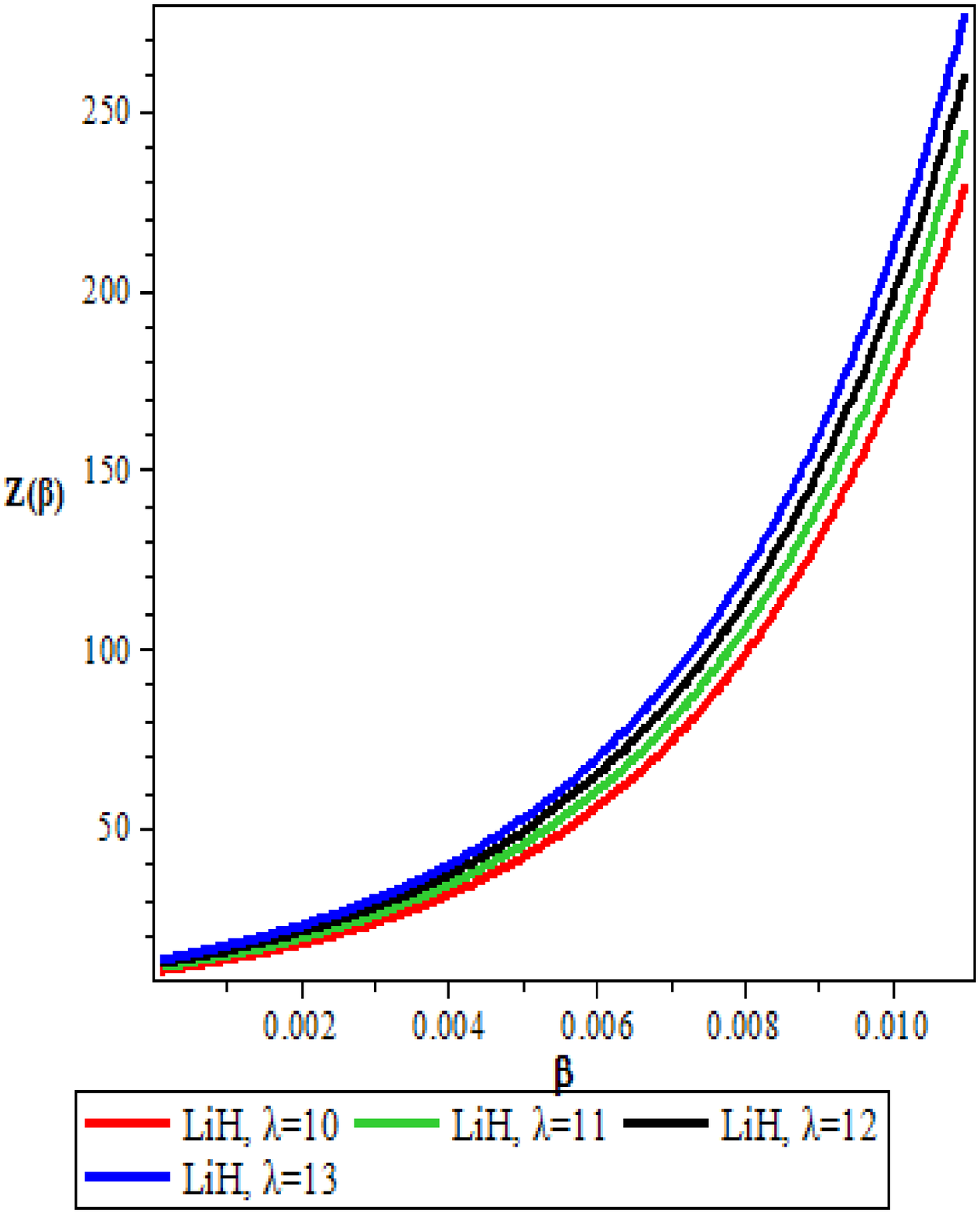}
		\caption*{Figure 5(a): Variation of Partition Function With Respect to $\beta$}
	\end{subfigure}\hspace{1cm} 
	~ 
	\begin{subfigure}[b]{0.5\textwidth}
		\includegraphics[width=9cm, height=9cm]{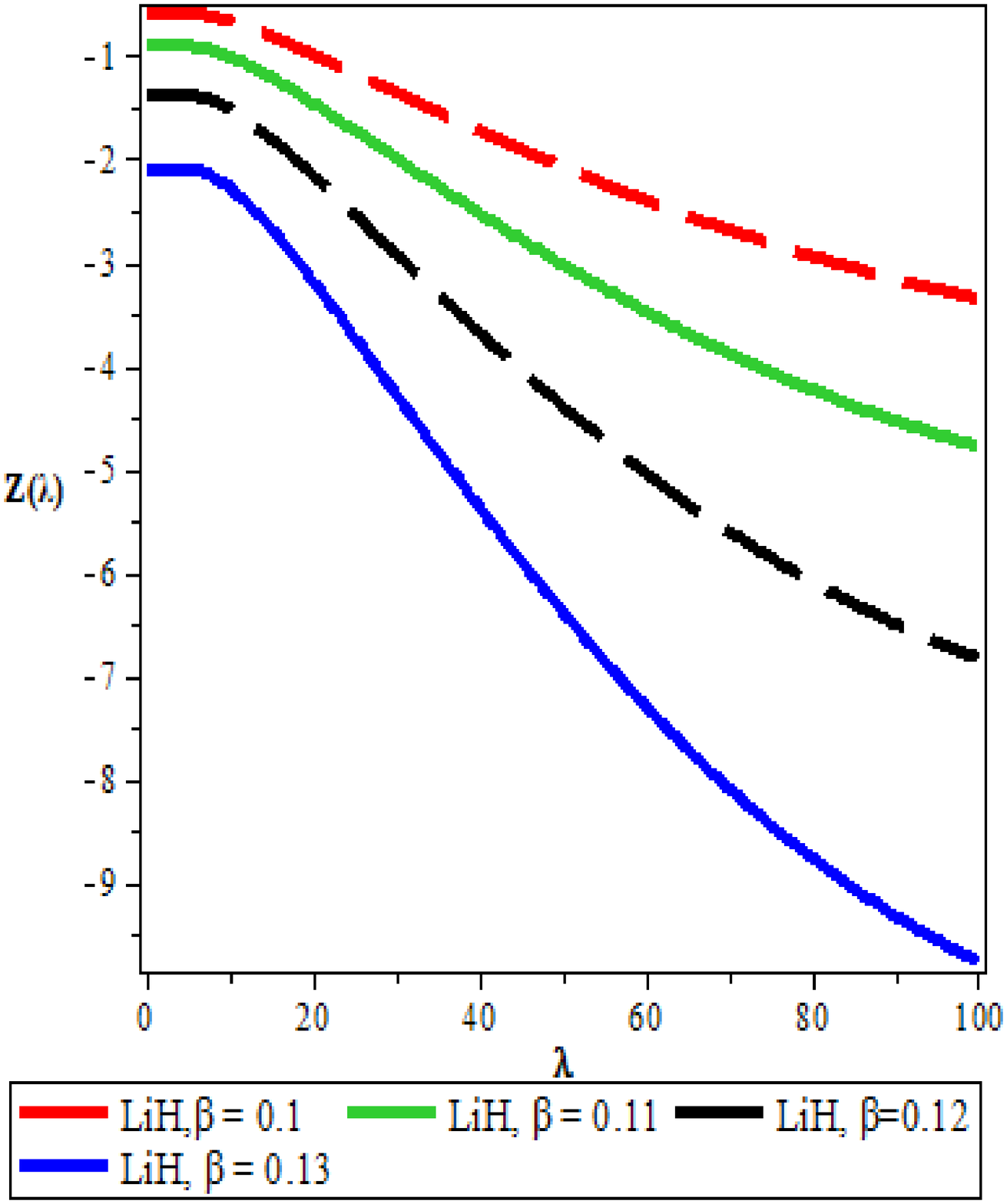}
		\caption*{Figure 5(b): Variation of Partition Function With Respect to $\lambda$}
	\end{subfigure}
	~ 
	\begin{subfigure}[b]{0.5\textwidth}
		\includegraphics[width=9cm, height=9cm]{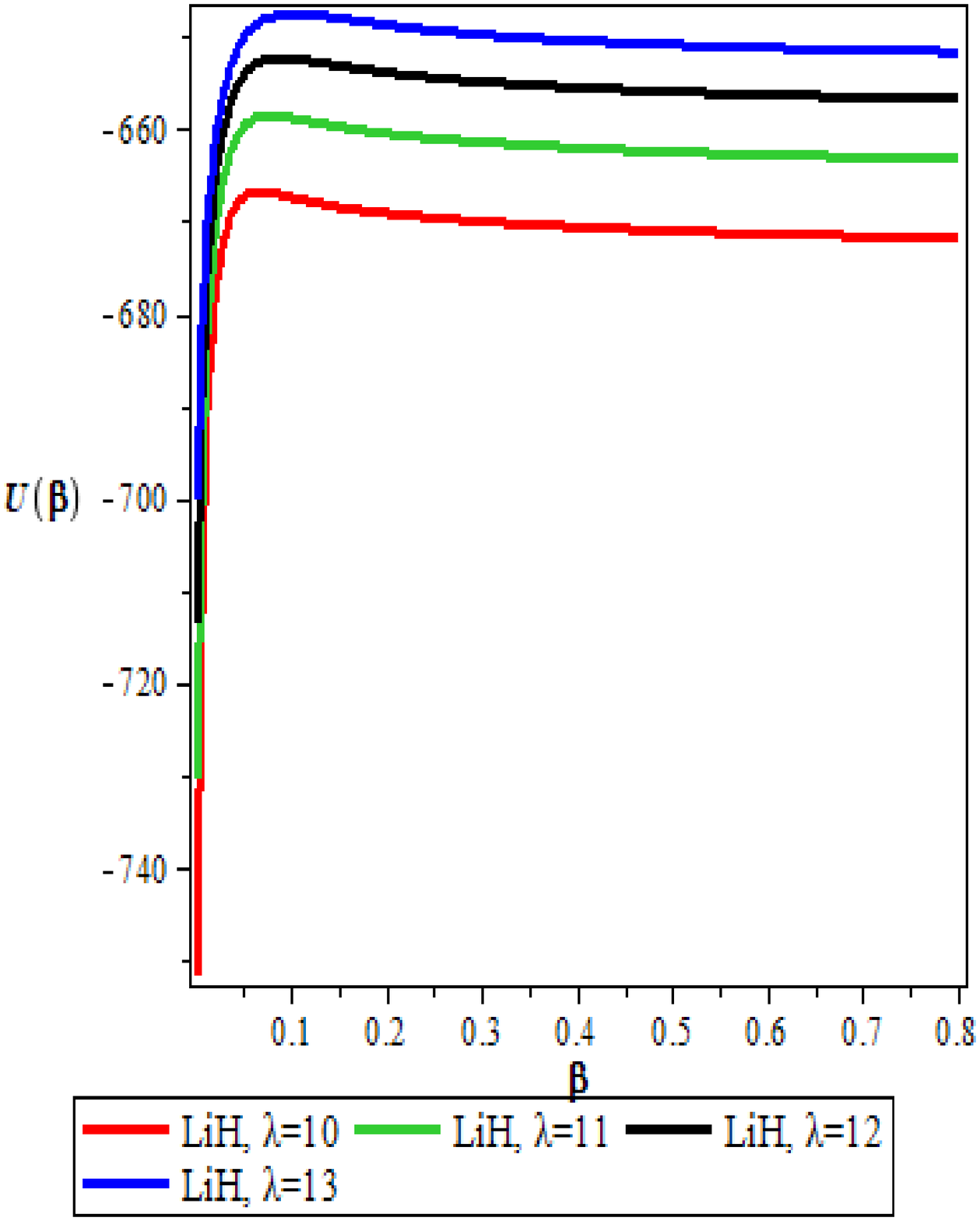}
		\caption*{Figure 6(a): Variation of Vibrational Mean Energy With Respect to $\beta$}
	\end{subfigure}\hspace{1cm}
	\begin{subfigure}[b]{0.5\textwidth}
		\includegraphics[width=9cm, height=9cm]{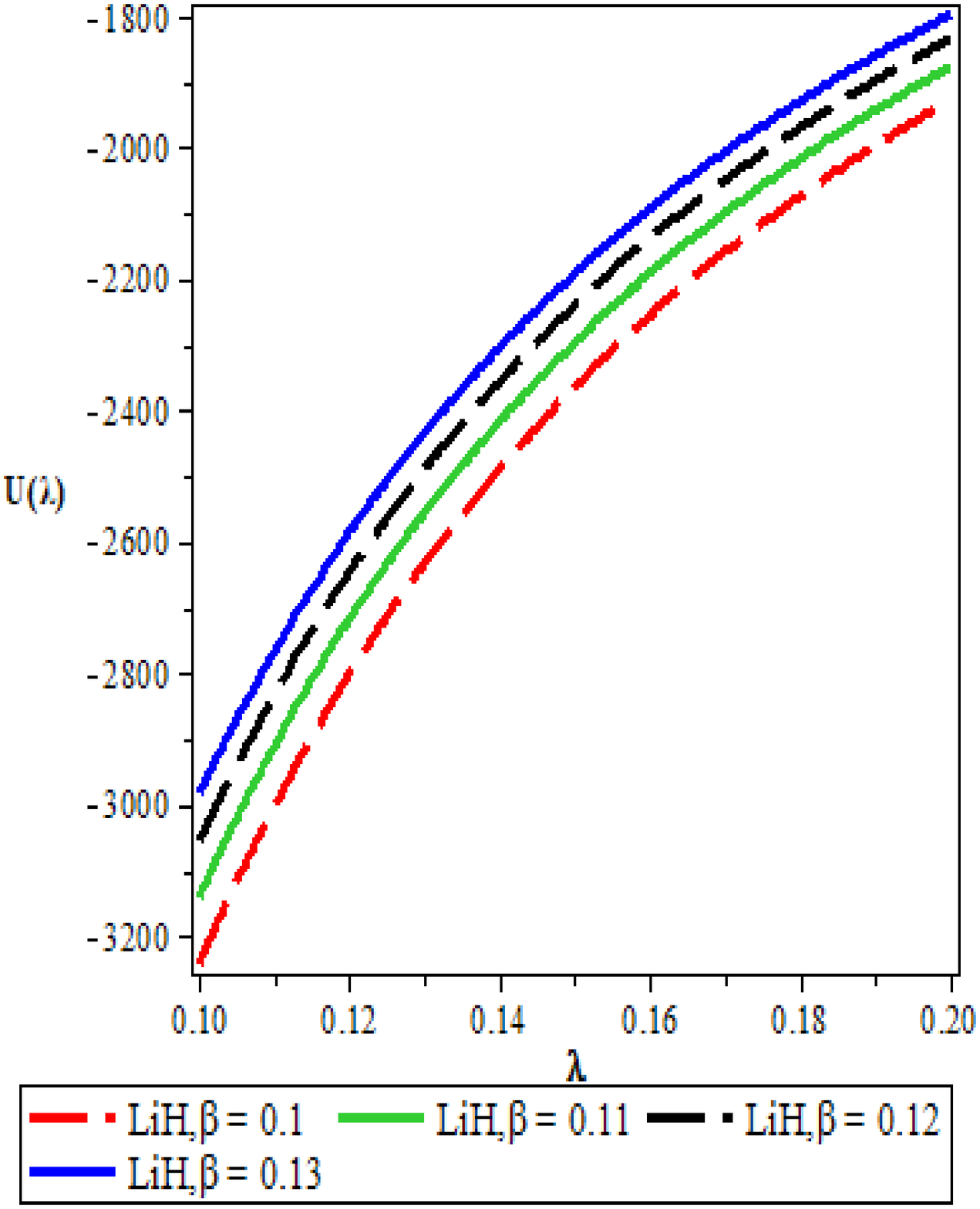}
		\caption*{Figure 6(b): Variation of Vibrational Mean Energy With Respect to $\lambda$}
	\end{subfigure}
\end{figure}
\begin{figure}
	\centering
	\begin{subfigure}[b]{0.5\textwidth}
		\includegraphics[width=9cm, height=9cm]{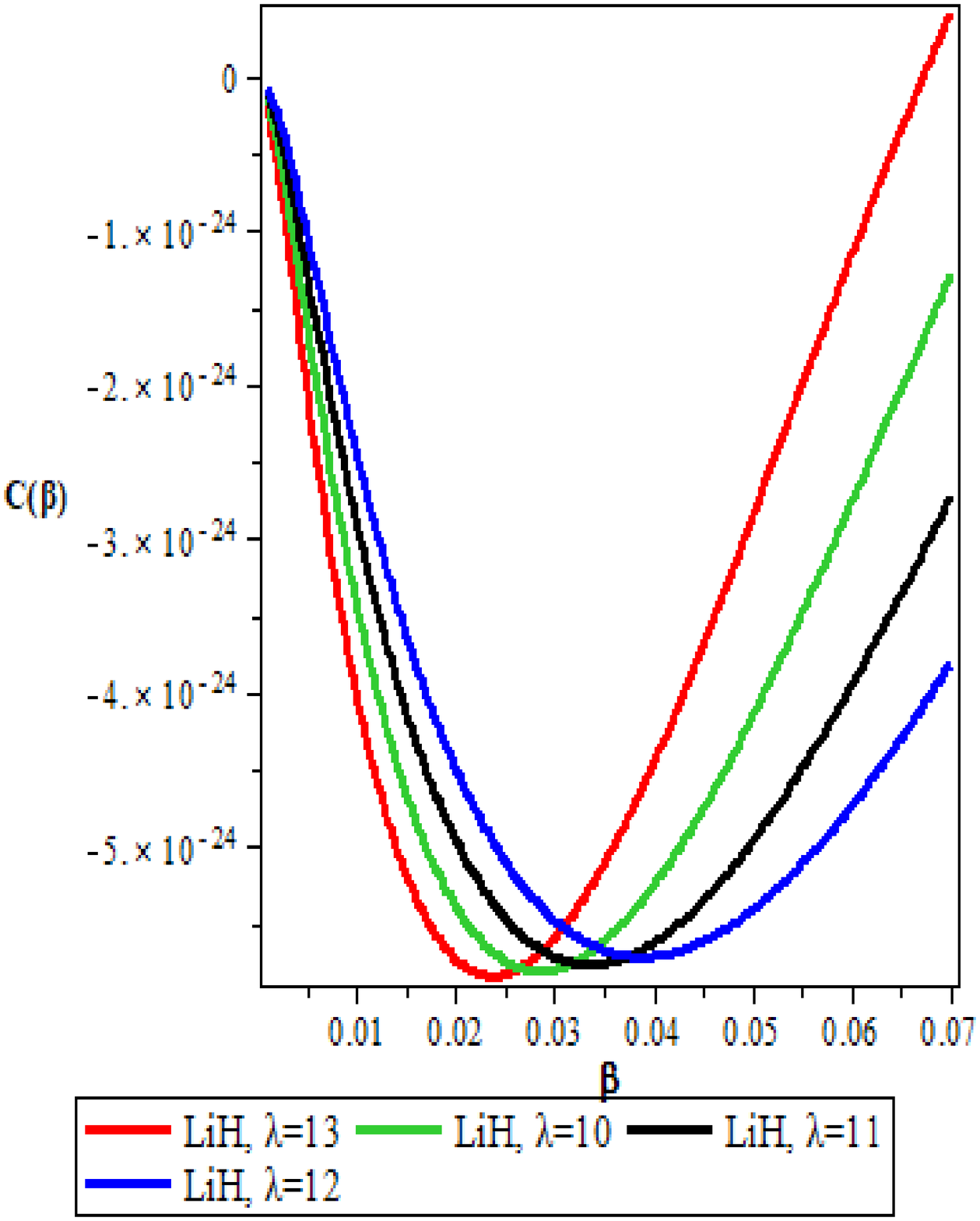}
		\caption*{Figure 7(a): Variation of Specific Heat Capacity With Respect to $\lambda$}
	\end{subfigure}
	~ 
	\begin{subfigure}[b]{0.5\textwidth}
		\includegraphics[width=9cm, height=9cm]{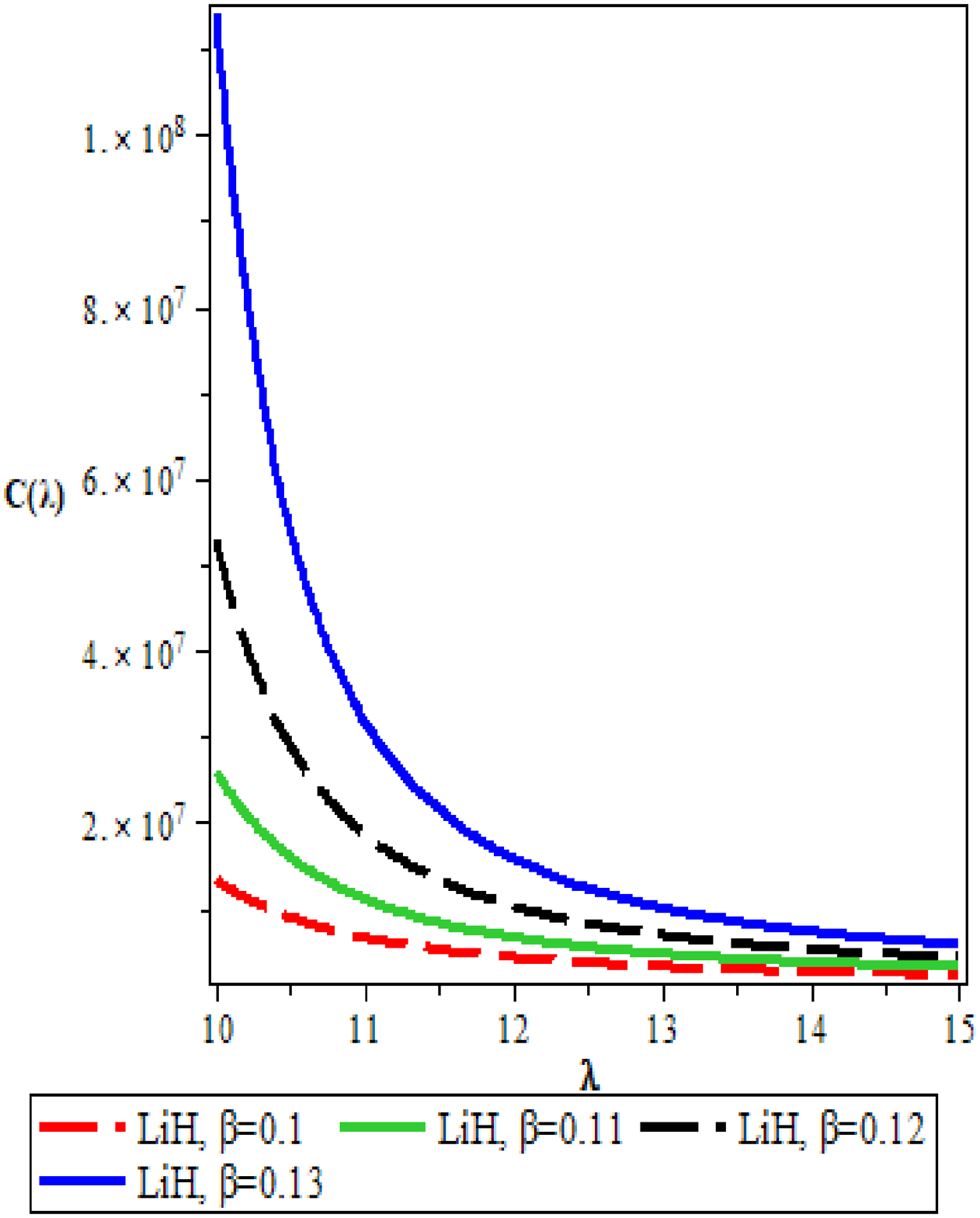}
		\caption*{Figure 7(b): Variation of Specific Heat Capacity With Respect to $\beta$}
	\end{subfigure}
	~ 
	\begin{subfigure}[b]{0.5\textwidth}
		\includegraphics[width=8cm, height=8cm]{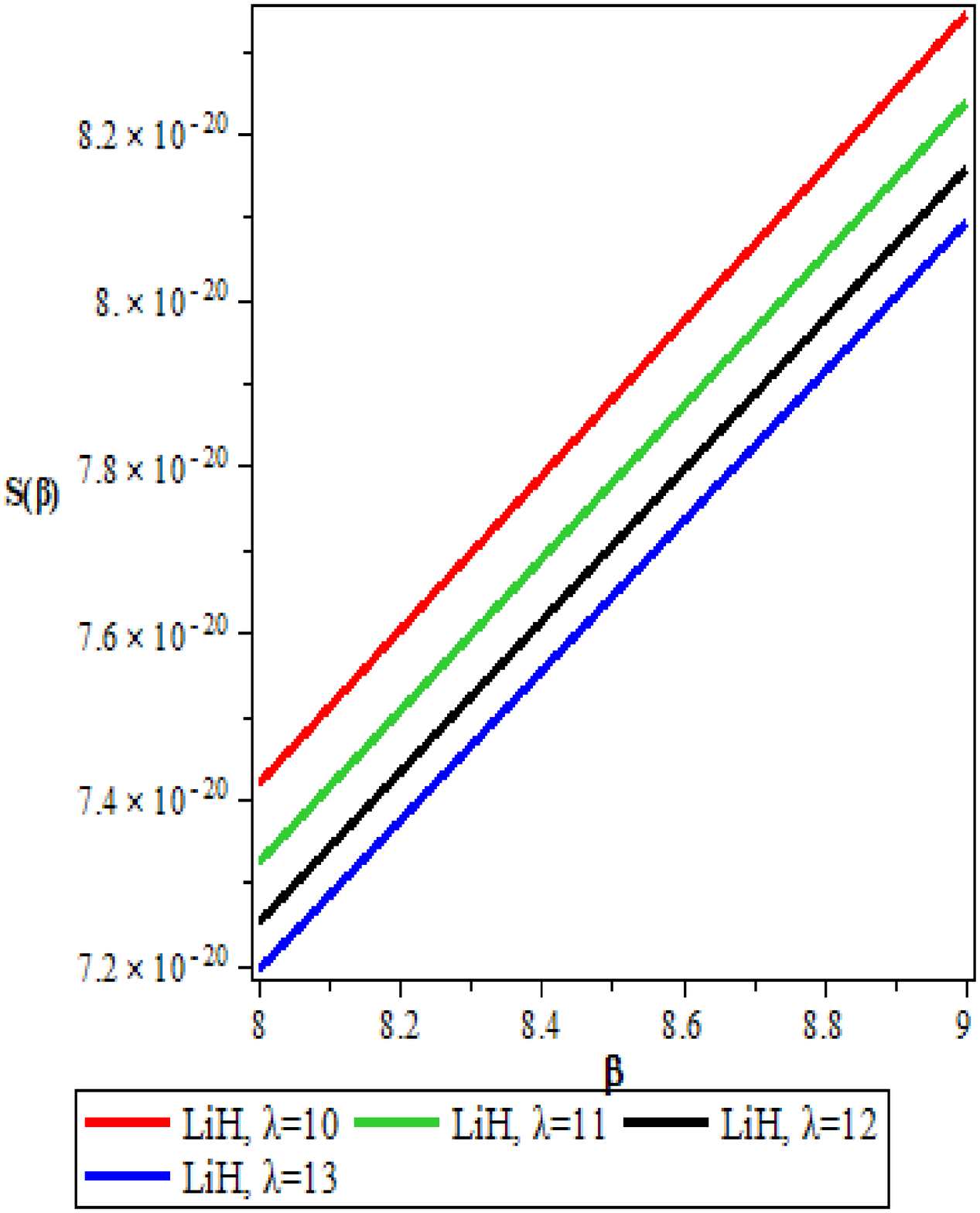}
		\caption*{Figure 8(a): Variation of Vibrational Entropy \\ With Respect to $\beta$}
	\end{subfigure}
	\begin{subfigure}[b]{0.5\textwidth}
		\includegraphics[width=8cm, height=8cm]{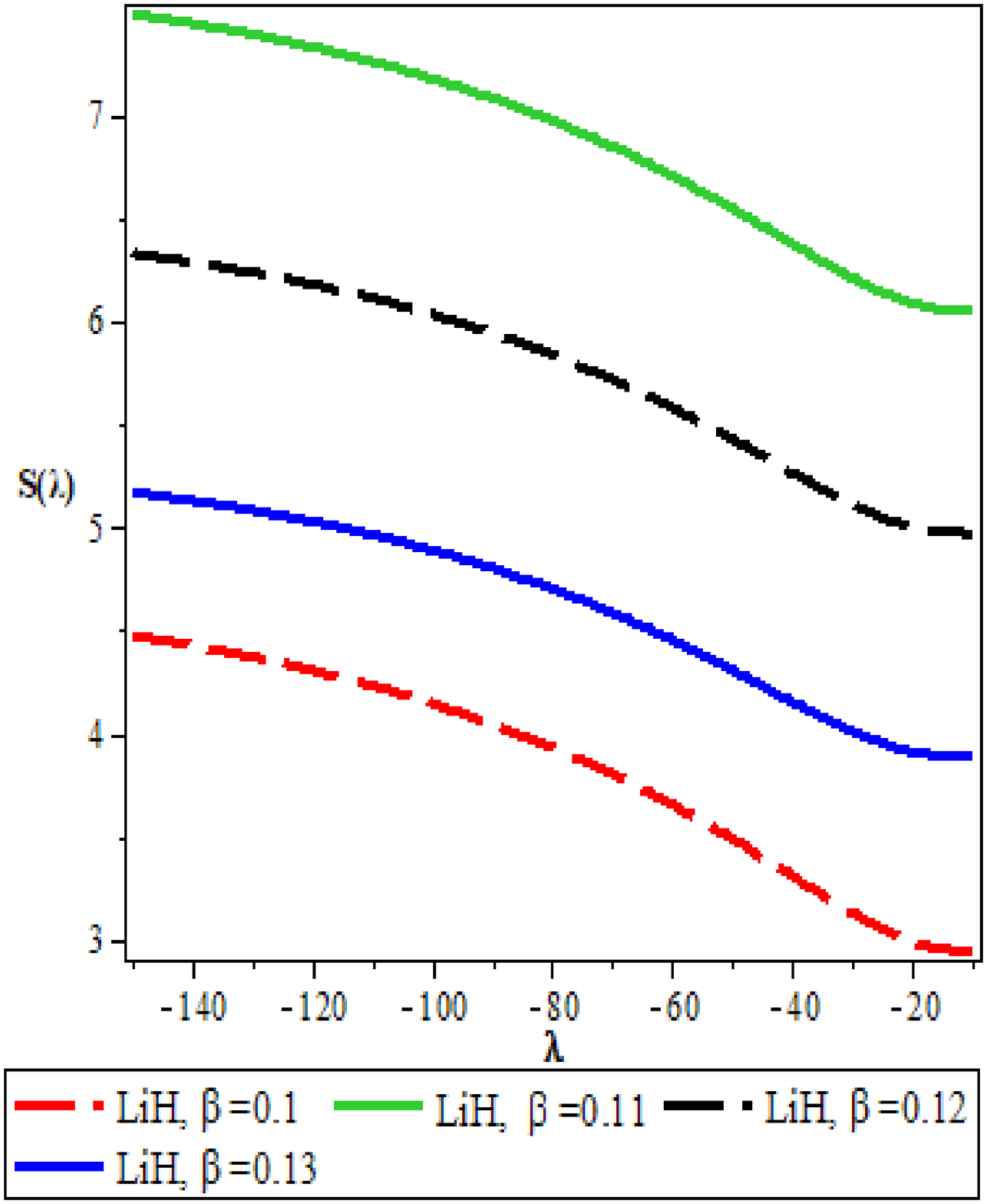}
		\caption*{Figure 8(b): Variation of Vibrational Entropy With Respect to $\lambda$}
	\end{subfigure}
	\caption*{}
\end{figure}

\begin{figure}
	\begin{subfigure}[b]{0.5\textwidth}
		\includegraphics[width=9cm, height=9cm]{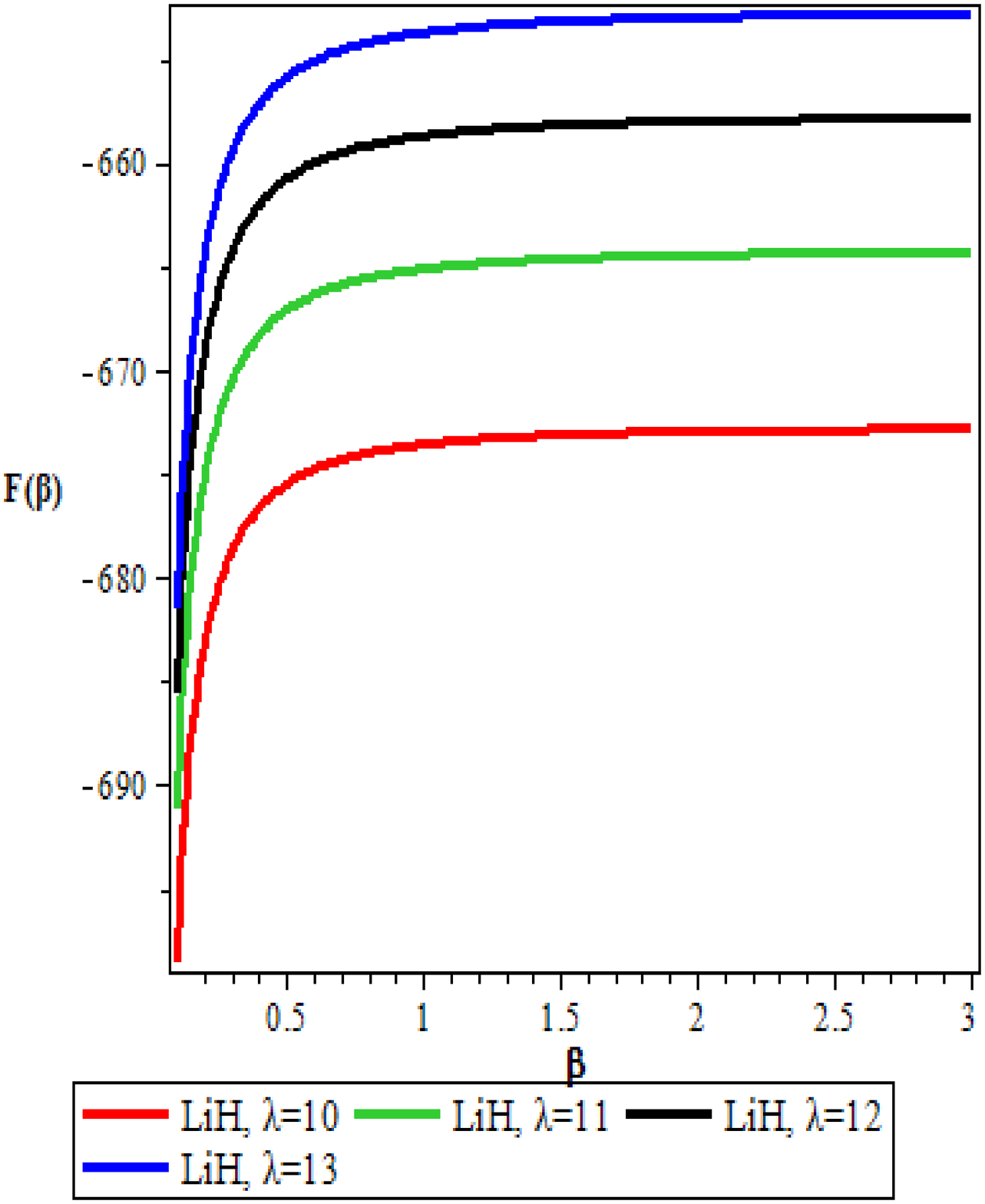}
		\caption*{Figure 9(a): Variation of Vibrational Free Energy With Respect to $\beta$}
	\end{subfigure}\hspace{1cm}
	~ 
\begin{subfigure}[b]{0.5\textwidth}
		\includegraphics[width=9cm, height=9cm]{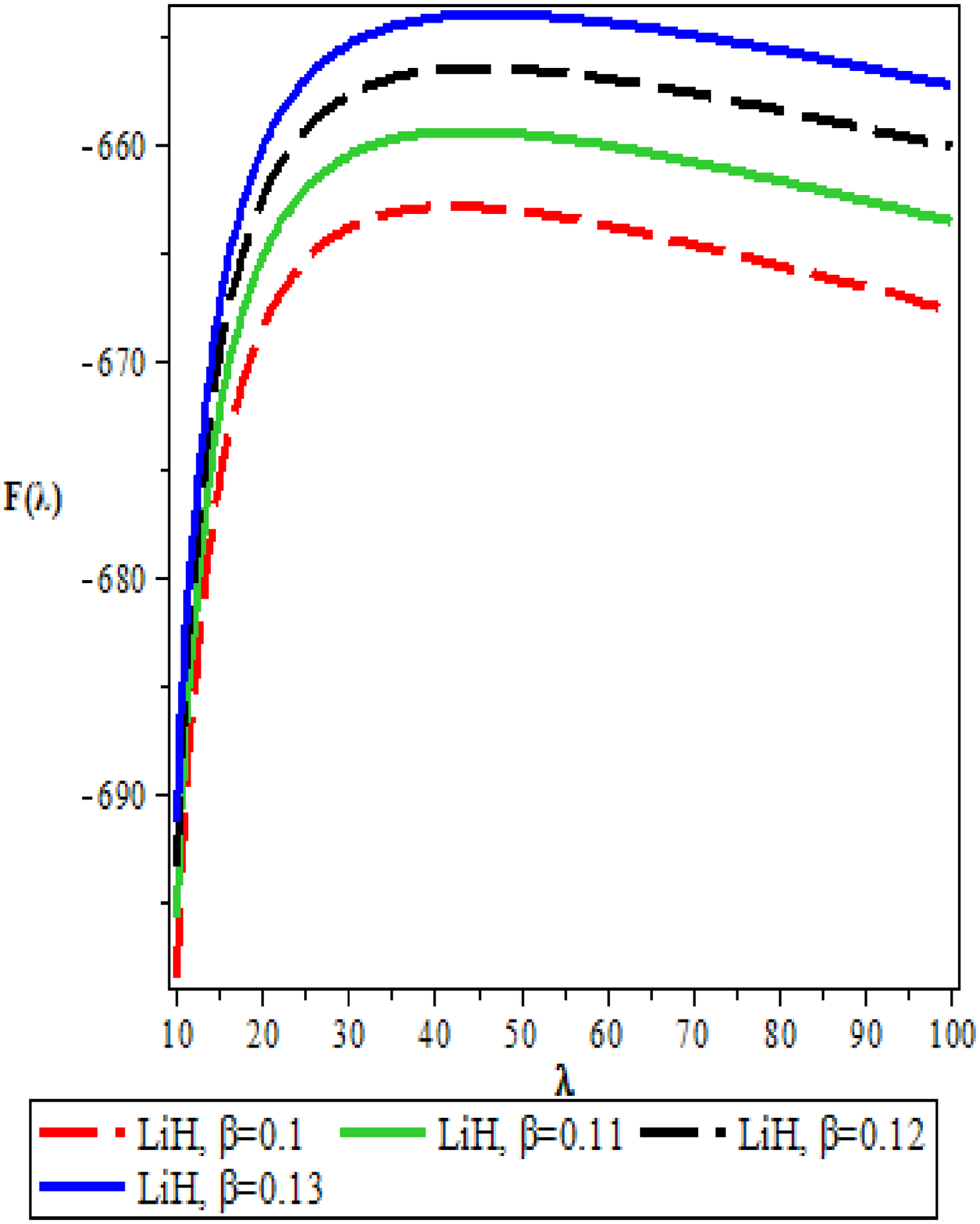} 
		\caption*{Figure 9(b): Variation of Vibrational Free Energy With Respect to $\lambda$}
	\end{subfigure}
\end{figure}

\section{Numerical Results and Discussion}
  The thermodynamics properties of KPGM was studied, the plots of the wavefunction and thermodynamics properties as a function of the inverse temperature parameter $\beta$ and $\lambda$ for Lithium hydride (LiH) diatomic molecule are shown in figures 1-4 and figures 5-9 respectively. Figure 1(a) is the wave function plot for fixed l=0 which begins with a commom origin and proceed to a continous sinusoidal curve with various maximum and minimum turning points for Lithium hydride molecule. From the graph, it can be observed that the peaks of the turning points increases with increase in the principal quantum number (n), as such n=0 has the lowest peak. The probability density curve for fixed l=0 is displayed in figure 1(b). This plot follows normal distribution with several maximum points which also increases with an increase in the principal quantum number. The probability density curve completely describe the localization of electrons of LiH molecule, hence electron is more localized at n=3 which has the highest maximum point. The same description of figure 1(a) is applicable to figures 2(a) and 4(a) while figures 2(b) and 4(b) also have the same description as figure 1(b). The wavefunction graph in figure 3(a) has maximum and minimum turning points at higher quantum states except at the ground state (n=0) where there is a divergence in the curve. In figure 3(b), the probability density plot is sinusoidal in nature which shows the localization of electrons for higher quantum state  while diverges at the ground state.

   Figures 5(a)and 5(b) show the variation of the vibrational partition function. It is observed in figure 5(a) that the partition function $Z(\beta)$ increases exponentially from the origin with increase in the inverse temperature parameter ($\beta$) but the partition function  $Z(\lambda)$ decreases with increase in $\lambda$ as presented in  figure 5(b)  for LiH diatomic molecule. The mean vibrational energy $U(\beta)$  as displayed in figures 6(a) increases monotonically with increase in the values of $\beta$ with slight maximum turning points.  The plot of $U(\lambda)$ against $\lambda$ has a hyperbolic nature. From figures 7(a), The vibrational specific heat capacity  $C(\beta)$  first decreases  with an increase in inverse temperature parameter to a minimum value and then increase monotonically. However,  the graph of vibrational specific heat capacity ($C(\beta)$) as a function of $\beta$ has various minimum turn points that touch the horizontal axis. $C(\lambda)$ decreases exponentially with $\lambda$ in figures 7(b). Plots of the vibrational entropy with different values of $\beta$ and $\lambda$ is shown in figures 8(a) and 8(b) respectively. As seen in figure 8(a), the vibrational entropy $C(\beta)$ increases linearly with increasing values of $\beta$ while  $S(\lambda)$ decreases with increasing values of $\lambda$. Plots of the mean free energy $F(\beta)$ as a function of $\beta$  increases monotonically with an increase in $\beta$  for various values of $\lambda$ as presented  in figure 9(a).   In figure 9 (b), the vibrational free energy exhibited an hyperborlic nature which increases with an increase in $\lambda$

\section{Conclusion}

In this work, we have solved the Schr\"{o}dinger wave equation in the presence of Kratzer plus generalized Morse potential (KPGM) using Parametric Nikiforov-Uvarov method. The energy eigenvalues and the corresponding wave function are obtained. However, we studied the thermodynamics properties of KPGM which are: vibrational partition Z, vibrational mean energy U, specific heat capacity C, entropy, and mean free energy. Also, we have plotted the variation of these thermodynamic functions as a function of $\beta$ and $\lambda$ for Lithium hydride diatomic molecule.

Conflict of interest---Not applicable
Funding: The article processing charge is completely funded by SCOAP3 and licensed under CCBY 4.0

\end{document}